\documentclass{article}
\usepackage{spconf,amsmath,amsfonts,graphicx,hyperref}
\usepackage{kotex}
\usepackage{booktabs}

\title{SPAM: Style Prompt Adherence Metric for Prompt-based TTS}
\name{Chanhee Cho\textsuperscript{\textdagger}, Nayeon Kim\textsuperscript{\textdagger}\thanks{\textsuperscript{\textdagger} Equal Contribution}, Bugeun Kim\thanks{This work was supported by the Institute of Information \& Communications Technology Planning \& Evaluation (IITP) grant funded by the Korea government (MSIT) [RS-2021-II211341, Artificial Intelligence Graduate School Program (Chung-Ang University)].}}
\address{\textit{Chung-Ang University}, \{cch991112, nayeonkim, bgnkim\}@cau.ac.kr}

\begin{document}
%
\maketitle
\begin{abstract}

Prompt-based text-to-speech (TTS) aims to generate speech that adheres to fine-grained style cues provided in a text prompt. However, most prior works depend on neither plausible nor faithful measures to evaluate prompt adherence. That is, they cannot ensure whether the evaluation is grounded on the prompt and is similar to a human. Thus, we present a new automatic metric, the Style Prompt Adherence Metric, which explicitly satisfies both plausibility and faithfulness. Inspired by the CLAP, our approach factorizes speech into acoustic attributes and aligns them with the style prompt. Also, we trained the scorer with a supervised contrastive loss, which could provide a clearer distinction between different semantics. We conducted two experiments on two perspectives. The plausibility experiment showed that SPAM achieved a strong correlation with the mean opinion score (MOS). Also, the faithfulness experiment demonstrated that SPAM is successfully grounded to the given style prompt, as it can discriminate different semantics of the prompt. We believe that SPAM can provide a viable automatic solution for evaluating style prompt adherence of synthesized speech.

\end{abstract}
\section{Introduction}
\label{sec:intro}

With the advancement of text-to-speech (TTS) technology, natural speech synthesis has been widely adopted in various applications. So, prompt-based TTS has gained traction to incorporate a richer set of stylistic cues as text prompt \cite{prompttts,promptstyle,prompttts++,voxinstruct,parlertts}. Nonetheless, automatic metrics that quantitatively assess how faithfully prompts are realized in synthesized speech remain underexplored. Most researches rely on Mean Opinion Score (MOS), which is labor and time-intensive. Thus, we propose a new automatic metric for prompt adherence.

Researchers have adopted various methods to assess how well synthesized speech adheres to the given style prompt automatically. Early researchers attempted to extract style embeddings and inspect whether these embeddings form a tight cluster of similar prompts \cite{prompttts, promptstyle, prompttts++}. However, it is questionable whether their evaluation appropriately mirrors human perception. As they depend on subjective visual inspection of cluster layouts, the distance between embeddings may not directly correspond to the perceptual distance. Thus, the method cannot be used for comparison between TTS models.

So, other researchers began to adopt the LLM-as-a-judge paradigm to obtain an automatic score of prompt adherence \cite{instructevaltts}. They let a large multimodal model assess the prompt adherence of a given prompt-speech pair. However, such a method cannot ensure whether judgment is truly grounded in the content of the style prompt. As LLMs are sensitive to small perturbations \cite{subtle_prompt1}, the result might not be \textit{faithful}. Here, in line with \cite{faithfulness}, we define an evaluation as \emph{faithful} when semantically similar prompts lead to consistent outcomes, while semantically distinct prompts yield divergent ones.

To tackle these two limitations of existing studies, we propose an approach inspired by CLAP \cite{clap}. Specifically, we posit that a contrastive-learning framework for computing text-audio similarity is well-suited to our problem.
Considering its potential, RA-CLAP \cite{ra-clap} adopted CLAP-style models to a task similar to ours, emotional speaking style retrieval. However, unlike that task, our task requires alignment of specific acoustic attributes. Yet, two challenges remain to apply CLAP methods.
First, factorization over acoustic attributes is required. Style prompts typically prescribe some acoustic attributes. As existing evaluators did not factorize attributes explicitly, it is questionable whether models actually consider such attributes during evaluation. Second, multi-positive examples may frequently occur within a single batch. Because of the birthday paradox, it is highly likely that a large batch contains multiple positive examples.
As standard CLAP loss does not consider this paradox, we need another loss function to exploit such a multi-positive situation.

Accordingly, we propose \textbf{S}tyle \textbf{P}rompt \textbf{A}dherence \textbf{M}etric, a new metric designed to overcome the above limitations. To enable quantitative automatic evaluation, we adopt a CLAP-based scorer that measures how well synthesized speech adheres to the given prompt. Also, to discriminate attributes through factorization, we design an SPAM that explicitly factors pitch, speed, energy, speaker, and transcription. Lastly, to utilize multi-positive examples, we adopt supervised contrastive (SupCon) loss \cite{supcon} instead of standard CLAP loss. Further, we tested the plausibility and faithfulness of SPAM.

\begin{figure*}[t]
    \centering
    \includegraphics[width=.95\linewidth]{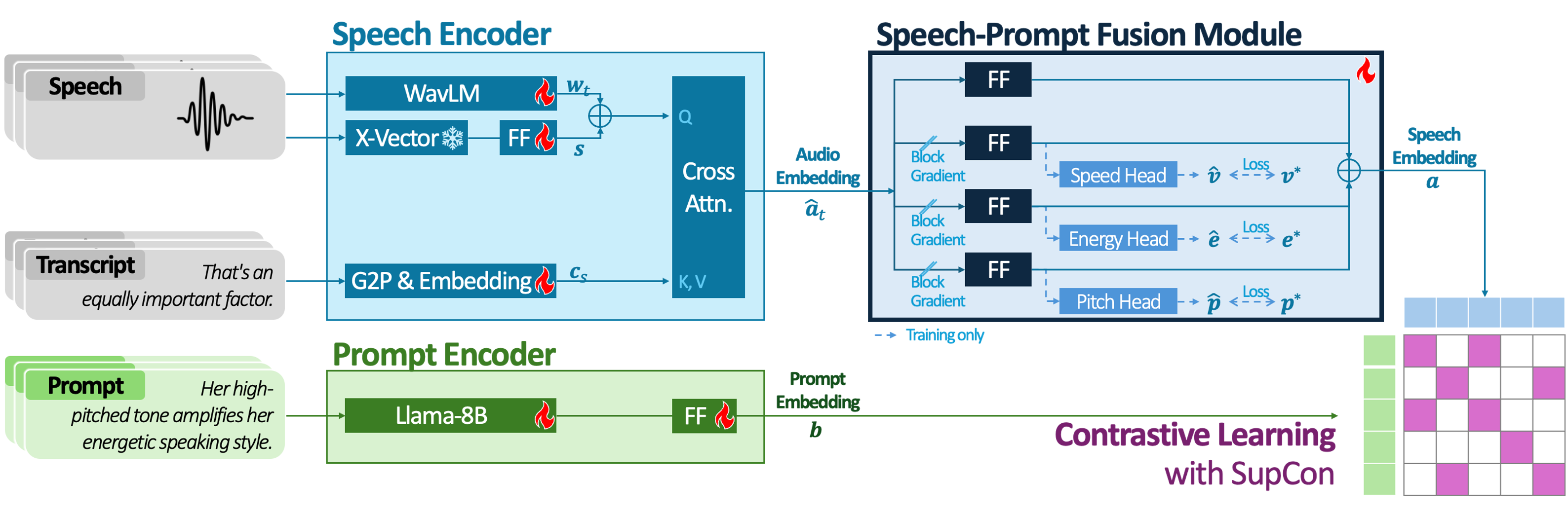}
    \caption{Architecture of SPAM}
    \label{fig:spam}
\end{figure*}

\section{The SPAM Metric}

SPAM can map audio and text into a common space and compute similarity between them. 
Specifically, we design SPAM to account for speech-specific factors and to output a similarity score via modules, as shown in Figure \ref{fig:spam}.
First, \textit{speech encoder} combines the waveform with other cues to produce an information-rich audio embedding. Second, \textit{prompt encoder} converts the text prompt into a prompt embedding. Third, \textit{speech-prompt fusion module} reinforces audio embedding with acoustics and computes similarity with the prompt.

\subsection{Speech Encoder}

To obtain a speech representation that is more fine-grained than the waveform itself, SPAM combines waveform with speaker and transcript. We adopt three encoders to encode those three. First, for waveform, we use WavLM \cite{wavlm} which showed reliable performance in many speech processing systems, including RA-CLAP \cite{ra-clap, Exploring-wavlm-on-speech-enhancement, wavllm}. Given a 16kHz waveform, the encoder produces frame-level embedding $\mathbf{w}_t\in \mathbb{R}^h$. Second, for speaker, we adopt X-Vector \cite{xvector}, a widely used embedding of speaker characteristics. We froze X-Vector module and adopted an feed-forward adapter that maps the x-vector output to $\mathbf{s} \in \mathbb{R}^h$. Third, for transcript, we directly embed phonemes of given transcript into $\mathbb{R}^h$ space. Specifically, we use a grapheme-to-phoneme (G2P) module and an embedding look-up table to obtain transcript embedding $\mathbf{c}_s$.

Next, we obtain audio embedding $\hat{\mathbf{a}}_t$ of given speech by combining $\mathbf{w}_t$, $\mathbf{s}$ and $\mathbf{c}_s$. We use a cross-attention layer to capture relation between waveform and transcript. Specifically, SPAM computes attention between $\mathbf{w}_t + \mathbf{s}$ vectors and $\mathbf{c}_s$ vectors. As a result, we obtain $\hat{\mathbf{a}}_t$ for each frame.

\subsection{Prompt Encoder}
To discriminate diverse text prompts, SPAM converts text prompt into prompt embedding $\mathbf{b}$ using a language model. The procedure is similar to CLAP or RA-CLAP except for the language model used. We adopt a Llama-3.1 8B \cite{llama-3.1} with a feed-forward adapter. As subtle differences in text prompt (e.g. tone, mood) might produce significant differences in the resulting speech, the prompt encoder should be large enough to be able to differentiate such differences. 

\subsection{Speech-Prompt Fusion Module}

To make SPAM recognize acoustic attributes separately, we input the audio embedding $\hat{\mathbf{a}}_t$ into four parallel branches: global waveform, speed, energy, and pitch. First, global waveform branch generates a global representation for overall waveform input. We adopt this branch to regularize training process and prevent over-fitting on a specific attribute.

The other three branches generate attribute-wise embedding. Through a feed-forward layer, each branch transforms $\hat{\mathbf{a}}_t$ into attribute-specific ones. Note that, to guide each acoustic branch to the corresponding signal, SPAM uses three auxiliary prediction heads; variance predictor from \cite{fastspeech2} for speed, and MLP heads for energy and pitch. These heads produce frame-level estimates $\hat{v}_t$, $\hat{e}_t$ and $\hat{p}_t$ for respectively.

After generating four embeddings, we obtain final speech embedding $\mathbf{a}$. To obtain an acoustically enhanced representation, we added all four embeddings for each frame and averaged the added result across frames. The two embeddings $\mathbf{a}$ and $\mathbf{b}$ are then passed to further computation. During the training, we compute contrastive and auxiliary losses. 
During the inference, we compute cosine similarity between them.

\subsection{Training Loss}

We compute total loss $\mathcal{L}$ using four losses during the training: contrastive $\mathcal{L}_{con}$, speed, energy, and pitch loss, as follows:
\begin{eqnarray*}
    \mathcal{L} &=& \lambda_{c} \mathcal{L}_{con} + \lambda_p \mathcal{L}_\delta(\hat{p}) +  \lambda_v \mathcal{L}_\delta(\hat{v}) +  \lambda_e \mathcal{L}_\delta(\hat{e}),\\
    \mathcal{L}_{\text{con}}
    &=&\frac{1}{2}\Big(
    \mathcal{L}^{\sup}(\mathbf{a}, \mathbf{b})
    +
    \mathcal{L}^{\sup}(\mathbf{b}, \mathbf{a})
    \Big)
\end{eqnarray*}
\begin{eqnarray*}
    \mathcal{L}^{\sup}(X, Y)
    &=&
    \sum_{\mathbf{x}\in X}
    \left(
    \mathbb{E}_{\mathbf{p}\sim P(\mathbf{x})}
    \left[
    -\log
    \frac{e^{\mathbf{x}^\top \mathbf{p}}}
         {\sum\limits_{\mathbf{y} \in Y} e^{\mathbf{x}^\top \mathbf{y}}}
    \right]
    \right),
\end{eqnarray*}
where $\mathcal{L}_{con}$ and $\mathcal{L}_\delta$ are SupCon loss \cite{supcon} and Huber loss, respectively. $\hat{p}$, $\hat{v}$, and $\hat{e}$ are utterance-level pitch, speech, and energy prediction by averaging $\hat{p}_t$, $\hat{v}_t$, and $\hat{e}_t$ values across frames. Also, $P(\mathbf{x})$ is a set of positive examples for anchor $\mathbf{x}$.

To define $P(\mathbf{a})$ and $P(\mathbf{b})$, SPAM use \textit{style key}. Popular prompt-based TTS datasets provide style keys \cite{textrolspeech, speechcraft, libritts-p}, which represent acoustic features of a speech audio; for example, a male voice with high pitch and normal speed. Thus, we exploit such key information to identify positive pairs; a prompt-audio pair is positive when the prompt and the audio map to the same style key. Note that, based on the style key, the matching problem between prompt and audio becomes a many-to-many problem; so, we used SupCon instead of InfoNCE to obtain better reliability during the training.

\section{EXPERIMENTS}

\begin{figure}
    \centering
    \includegraphics[width=\linewidth]{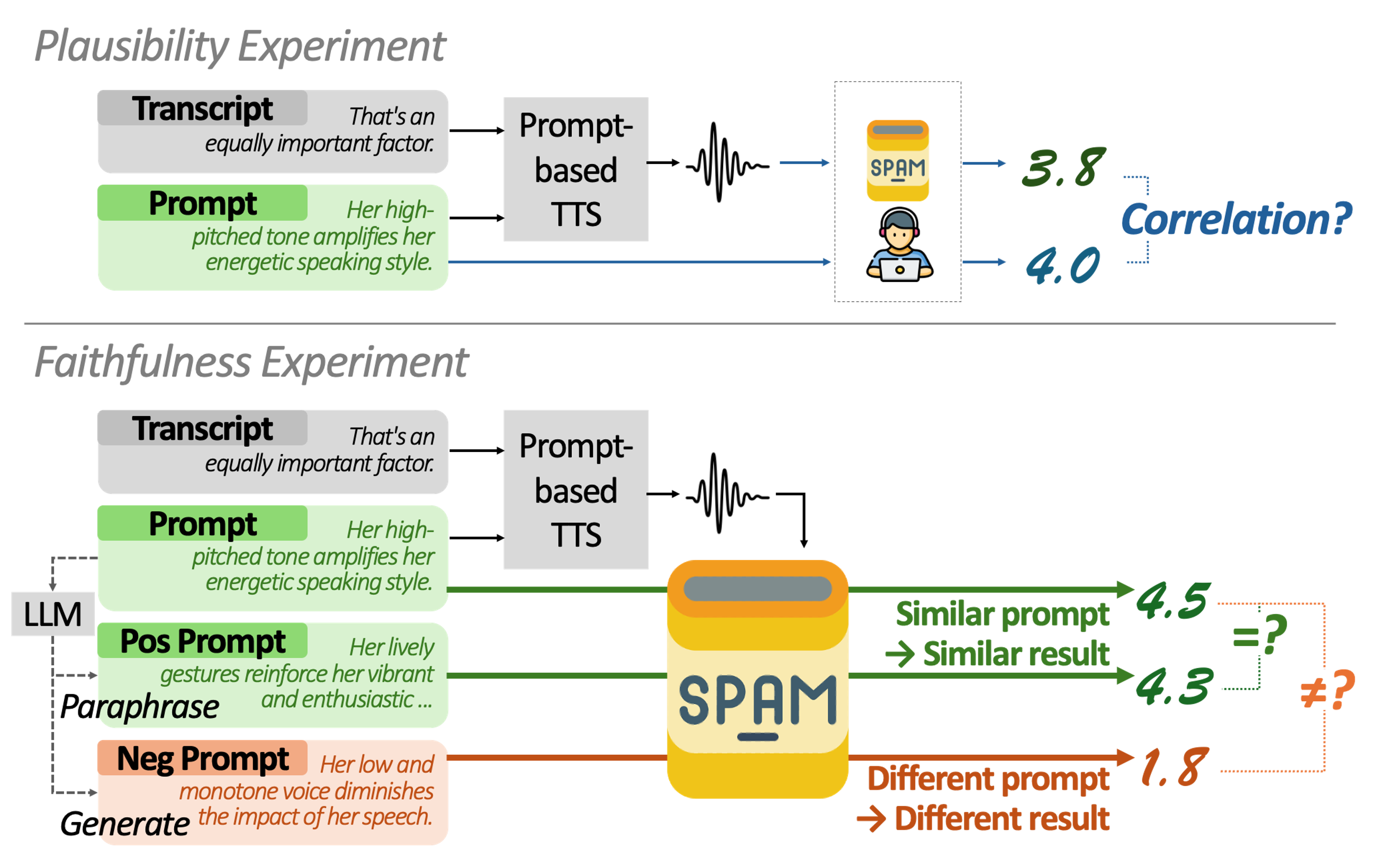}
    \caption{Plausibility and Faithfulness Experiment}
    \label{fig:experiment}
\end{figure}

To ensure the validity of our SPAM metric, we conducted two experiments (see Figure \ref{fig:experiment}). First, \textit{plausibility} experiment examines whether a metric successfully evaluates prompt adherence as human does. Second, \textit{faithfulness} experiment examines whether a metric clearly discriminates prompt-adhering speech from the opposite one.

\subsection{Plausibility experiment}

To assess plausibility, we compared metrics with human judgments of prompt adherence. Specifically, we measured the Mean Opinion Score (MOS) by recruiting 320 annotators through CloudResearch, who rated adherence on a 5-point Likert scale. For each prompt–speech pair, 8 annotators provided ratings, which were averaged to obtain MOS.

Using utterance-level correlation, we evaluate the plausibility of metric evaluation. When a correlation between MOS and a metric is high enough, we could conclude that the metric mirrors human evaluation well. We used three measures, Linear Correlation Coefficient (LCC), Spearman Rank Correlation Coefficient (SRCC), and Kendall’s Tau (KTAU).

\subsection{Faithfulness experiment}

Also, to assess faithfulness, we compare the responses of metrics when we provide a variation of aㅜ original prompt. A faithful metric should satisfy two conditions: (1) The score distributions should be the same when we input the original prompt attached to the speech and a semantically equivalent paraphrase; and (2) the score distribution should be left-shifted relative to that of the original prompt when evaluating semantically inequivalent prompts. In this paper, we call the former prompt \textit{positive prompts} and the latter \textit{negative prompts}. For each prompt-speech pair, we generated 10 positive and 10 negative prompts and compared how metrics react differently to original, positive, and negative prompts.

We evaluated faithfulness in two ways: adherence rate (AR) and paired $t$-test. First, AR measures whether positive prompts received a higher score than negative ones. Similar to a bootstrapping method, we calculated AR as average probability of such cases on all possible pairs of positive-negative prompts. Second, the paired $t$-test measures whether two conditions are satisfied. We tested the following alternative hypotheses. \\(1) $H_{1}$: $\mu_+ \ne \mu_0$ indicating positive prompt ($\mu_+$) was scored unequal to original ($\mu_0$). A good metric should reject $H_1$. \\(2) $H_2$: $\mu_- < \mu_0$ indicating negative prompt ($\mu_-$) received scores lower than original. A good metric should accept $H_2$.

\newcommand{\starp}{\textsuperscript{*}}
\newcommand{\starpp}{\textsuperscript{**}}
\newcommand{\starppp}{\textsuperscript{***}}
\begin{table*}[!t]
    \centering
    \small
    \begin{tabular}{l|r@{\;\;}r@{\;\;}r|r|r@{$\pm$}r|r@{$\pm$}r@{\;\;}r@{}l@{\;}c|r@{$\pm$}r@{\;\;}r@{}l@{\;}c}
        \toprule
         & \multicolumn{3}{c|}{\textit{Plausibility}} & \multicolumn{13}{c}{\textit{Faithfulness}}\\
        \cmidrule(lr){2-4}
        \cmidrule(lr){5-17}
         & LCC & SRCC & KTAU & AR & \multicolumn{2}{c|}{Original} & \multicolumn{2}{c}{Positive} & \multicolumn{3}{c|}{($H_1$ reject?)} & \multicolumn{2}{c}{Negative}  & \multicolumn{3}{c}{($H_2$ accept?)}\\
        \midrule
        \multicolumn{17}{l}{On \textit{TextrolSpeech} dataset}\\
        \midrule
        SPAM (WavLM) & 0.584 & 0.584 & 0.405 & \textbf{0.862} & 0.361 & 0.153 &
                       0.357 & 0.143 & \textbf{-2.025} &  & \checkmark & 0.050 & 0.221 & -\textbf{20.145} & \starppp & \checkmark \\
                       
        \phantom{SPAM} (CLAP)  & 0.554 & 0.560 & 0.389 & 0.841 & 0.039 & 0.026 &
                       0.035 & 0.025 & -3.699 &\starppp & & -0.005 & 0.030 & \textbf{-17.538} & \starppp & \checkmark \\
                       
        RA-CLAP      & 0.520 & 0.514 & 0.357 & 0.852 & 0.400 & 0.324 &
                       0.380 & 0.312 & -3.479 &\starpp & & -0.020 & 0.219 & \textbf{-16.912} & \starppp & \checkmark \\
        \midrule
        \multicolumn{17}{l}{On \textit{LibriTTS-P} dataset}\\
        \midrule
        SPAM (WavLM) & 0.580 & 0.568 & 0.400 & \textbf{0.771} & 0.279 & 0.171 &
                       0.294 & 0.162 & 3.200 &\starpp & & 0.052 & 0.234 & \textbf{-15.105} & \starppp & \checkmark\\
                       
        \phantom{SPAM} (CLAP)  & 0.516 & 0.499 & 0.346 & 0.766 & 0.027 & 0.027 &
                       0.028 & 0.027 & \textbf{0.239} &  &\checkmark & -0.006 & 0.033 & \textbf{-14.689} & \starppp & \checkmark\\
                       
        RA-CLAP      & 0.429 & 0.435 & 0.304 & 0.750 & 0.249 & 0.285 &
                       0.284 & 0.297 & 2.645 &\starp & & 0.003 & 0.199 & \textbf{-7.628} & \starppp & \checkmark \\
        \bottomrule
        \multicolumn{17}{r}{\starp $p<0.05$, \starpp $p<0.01$, \starppp $p<0.001$}
    \end{tabular}
    \caption{Result of plausibility and faithfulness experiment, on the entire test set}
    \label{tab:all}
\end{table*}

\begin{table}[!t]
    \centering
    \begin{tabular}{l|ccc}
        \toprule
         & \multicolumn{2}{c}{SPAM} & RA- \\
         \cmidrule(lr){2-3}
         & (WavLM) & (CLAP) & CLAP\\
        \midrule
        \multicolumn{4}{l}{On \textit{TextrolSpeech} dataset}\\
        \midrule
        Ground truth audio & 0.721 & 0.673 & \textbf{0.726} \\
        ParlerTTS-mini & \textbf{0.749} & 0.710 & 0.709 \\
        \phantom{ParlerTTS}-large & 0.648 & \textbf{0.653} & 0.615 \\
        PromptTTS   & \textbf{0.665} & 0.584 & 0.416 \\
        PromptStyle & 0.515 & \textbf{0.518} & 0.503 \\
        VoxInstruct & 0.462 & 0.429 & \textbf{0.473} \\
        \midrule
        \multicolumn{4}{l}{On \textit{LibriTTS-P} dataset}\\
        \midrule
        Ground truth audio & \textbf{0.718} & 0.667 & 0.545 \\
        ParlerTTS-mini & \textbf{0.752} & 0.712 & 0.536 \\
        \phantom{ParlerTTS}-large & \textbf{0.716} & 0.610 & 0.531 \\
        PromptTTS   & \textbf{0.560} & 0.495 & 0.375 \\
        PromptStyle & \textbf{0.519} & 0.503 & 0.417 \\
        VoxInstruct & \textbf{0.454} & 0.403 & 0.371 \\
        \bottomrule
    \end{tabular}
    \caption{Linear correlation coefficient between human MOS and automatic metrics, for each model and dataset}
    \label{tab:dataset}
\end{table}

\subsection{Datasets and Baselines}

For training SPAM, we used the combined train sets of TextrolSpeech \cite{textrolspeech} and SpeechCraft \cite{speechcraft}, using only ground-truth data with high-quality voices.

For the experiment, we used two test sets: TextrolSpeech and LibriTTS-P [20], the latter used to verify prompt adherence on unseen styles. We randomly sampled 50 prompt–speech pairs from each dataset.

Furthermore, we generated five synthesized speeches for each ground-truth speech, using five prompt-based TTS: PromptTTS \cite{prompttts}, PromptStyle \cite{promptstyle}, VoxInstruct \cite{voxinstruct}, Parler-TTS-mini-v1, and Parler-TTS-large-v1 \cite{parlertts}. This produced 100 ground truth and 500 synthesized speech for each experiment. Also, we measured MOS with 600 original and 600 negative pairs to include low adherence cases.

Using the dataset, we compare two versions of SPAM with RA-CLAP teacher model \cite{ra-clap}. Two variants of SPAM use different waveform encoders: WavLM and CLAP encoder. RA-CLAP is slightly different from ours as it is designed emotional speaking style retrieval. Also, RA-CLAP discriminate acoustic attributes, while SPAM can.

\section{RESULT AND DISCUSSION}

\textbf{Plausibility.}
The result of the plausibility experiment is shown in Table \ref{tab:all} (left). In general, we observed moderate correlation between MOS and all metrics, regardless of the correlation coefficient. Specifically, SPAM (WavLM) showed LCC around 0.58, which is higher than that of RA-CLAP (0.520 and 0.429). Furthermore, when we analyze per-model coefficient as shown in Table \ref{tab:dataset}, we observed stronger and stable correlation between SPAM and MOS. For example, SPAM showed strong stable correlation (around 0.72) on ground truth audio for each dataset. Meanwhile, RA-CLAP showed a large gap across datasets: 0.726 and 0.545. This phenomenon is constantly observed across different TTS models (e.g., ParlerTTS). Such consistency of SPAM reveals its generalizability towards unseen style prompts. We suspect two possible causes of such consistency: discrimination of acoustic attributes via the speech-prompt fusion module, and enhancement of context understanding via the Llama 3 prompt encoder. Because of these two, SPAM can replace MOS for evaluating prompt adherence since it provides stable and plausible scoring regardless of prompt style.

\textbf{Faithfulness.}
The result of the faithfulness experiment is shown in Table \ref{tab:all} (right). Regarding AR, SPAM (WavLM) showed a higher AR score (0.862 and 0.771) than RA-CLAP (0.852 and 0.750). Such a high AR score indicates that SPAM can successfully discriminate positive and negative examples by assigning a higher score to positive prompts than to negative prompts. Results from a paired $t$-test provide a more rigorous explanation. The test revealed that only the SPAM series almost successfully scored positive examples with equal scores to the original, as the alternative hypothesis ($H_1$) is rejected. Meanwhile, RA-CLAP failed to reject the hypotheses, indicating unfaithful cases that positive examples can have unequal score to the original one. For the negative cases where contrastive loss aims to learn, all metrics successfully passed the test by accepting the alternative hypothesis ($H_2$). We suspect that this difference in positive examples is due to the adoption of SupCon loss, which can utilize the relationship between positive examples more than standard CLAP loss. Thus, SPAM demonstrated faithful evaluation.

\section{Conclusion}

In this paper, we proposed \textbf{S}tyle \textbf{P}rompt \textbf{A}dherence \textbf{M}etric for evaluating whether synthesized speech adheres to the prompt. We modified CLAP to satisfy two requirements of the task: alignment of acoustic attributes and consideration of multi-positive samples. Through plausibility and faithfulness experiments, we evaluated whether SPAM produces similar outcomes to humans and discriminates positive and negative prompts. As a result, we demonstrated that SPAM provides a plausible and faithful evaluation for prompt adherence.

\newpage
\bibliographystyle{IEEEbib}
\bibliography{strings,refs}

\end{document}